\documentclass[twocolumn,prl,,showpacs,floatfix]{revtex4-1}
\usepackage{graphicx}
\usepackage{amsmath}
\usepackage{bm}
\usepackage{epsfig}
\usepackage{amsfonts}
\usepackage{amssymb}
\usepackage[usenames]{color}
\begin{document}

\newcommand{\brm}[1]{\bm{{\rm #1}}}
\newcommand{\tens}[1]{\underline{\underline{#1}}}
\newcommand{\muk}{\mu}
\newcommand{\kappat}{{\tilde \kappa}}

\newcommand{\Ochange}[1]{{\color{red}{#1}}}
\newcommand{\Ocomment}[1]{{\color{PineGreen}{#1}}}
\newcommand{\Oremove}[1]{{\color{Green}{}}}
\newcommand{\Tcomment}[1]{{\color{Blue}{#1}}}
\newcommand{\Tchange}[1]{{\color{ProcessBlue}{#1}}}
\newcommand{\Tremove}[1]{{\color{Purple}{}}}
\title{Simple lattice model for biological gels}
\author{Olaf Stenull}
\affiliation{Department of Physics and Astronomy, University of
Pennsylvania, Philadelphia PA 19104, USA}
\author{T. C. Lubensky}
\affiliation{Department of Physics and Astronomy, University of
Pennsylvania, Philadelphia PA 19104, USA}

\date{\today}

\begin{abstract}
We construct a three-dimensional lattice model for biological
gels in which straight lines of bonds correspond to filamentous
semi-flexible polymers and lattice sites, which are exactly four-fold
coordinated, to crosslinks.  With only stretching central forces
between nearest neighbors, this lattice is sub-isostatic with an
extensive number of zero modes; but all of its elastic constants are
nonzero, and its elastic response is affine. Removal of bonds with
probability $1-p$ leads to a lattice with average coordination number
less than four and a distribution of polymer lengths. When bending
forces are added, the diluted lattice exhibits a rigidity threshold at
$p=p_b<1$ and crossover from bending-dominated nonaffine to
stretching-dominated affine response between $p_b$ and $p=1$.
\end{abstract}
\pacs{87.16.Ka, 	
62.20.de, 	
61.43.-j, 	
05.70.Jk 	
}

\maketitle

Biological gels~\cite{Alberts,Elson1988,JanmeySto1990,Kasza} are
elastic networks, formed by crosslinked semiflexible polymers, that
play a critical role in determining and controlling the mechanical
properties of eukaryotic cells. Here we introduce and analyze
properties of a three-dimensional (3d) lattice model for these gels in which
straight sequences of bonds correspond to polymers, and lattice sites,
with a maximum coordination number of four, correspond to crosslinks.

\begin{figure}
\centerline{\includegraphics{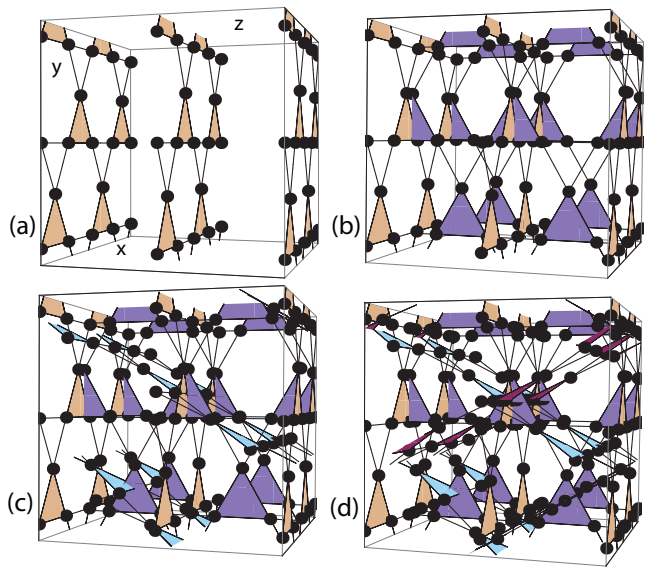}}
\caption{(Color online) Steps leading from a stack of kagome lattices (a) to our final model lattice (d).}
\label{fig:construction}
\end{figure}

Much of our intuition about filamentous networks comes from studies of
two-dimensional (2d) Mikado models
\cite{HeadMacKLevine2003,WilhelmFrey2003} in which straight lines,
representing semi-flexible polymers of length $L$ with stretching
modulus $\muk$ and bending modulus $\kappa$, are laid down randomly on
a plane and crosslinked at their crossing points. These studies show
that there is a crossover from non-affine, bending dominated to affine,
stretching dominated response as the number of crosslinks per polymer
is increased~\cite{HeadMacKLevine2003,WilhelmFrey2003,OnckGiessen2005}.
The kagome lattice [Fig.~\ref{fig:construction}a] with coordination
number $z=4$ is a periodic version of the infinite $L$ limit of the
Mikado model, albeit with a monodisperse distribution of lengths
between neighboring crosslinks (segment lengths). This lattice with
nearest-neighbor springs only and no bending energy exhibits a
nonvanishing shear modulus \cite{SouslovLub2009} and affine response
even though it is just on the verge of mechanical instability: With
$z=2d$ ( where $d$ is the spatial dimension) under periodic conditions,
it is exactly isostatic \cite{Maxwell1864,Calladine1977}, but it has a
number of zero modes that scales as its perimeter
\cite{SouslovLub2009}.  It provides a rigorous demonstration of the
existence of a lattice with an $L\to \infty$ affine limit such as seen
in simulations on the random
lattice~\cite{HeadMacKLevine2003,WilhelmFrey2003}. A network of
crosslinked semi-flexible polymers in 3d still has a maximum
of only four neighbors per crosslink, and it is subisostatic in the
absence of bending forces with a number of zero modes that scales with
its volume. It is, therefore, not obvious that the affine,
stretching-dominated shear-rigid limit found in 2d can exist in such
networks even though models assuming affine response are in
good agreement with experimental
measurements~\cite{Storm&Co2005}. Indeed, 3d
computer-generated filamentous networks \cite{Huisman1,HuismanLub2010}
show bending but not stretching dominated linear response.

\begin{figure}
\centerline{\includegraphics[width=5.8cm]{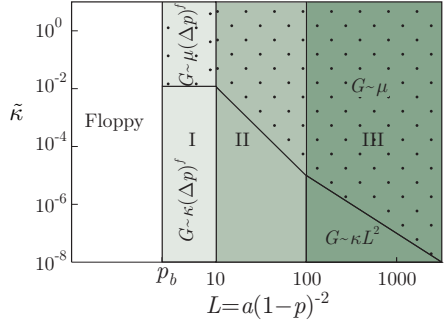}}
\caption{(Color online) Schematic phase diagram for the $3d$ kagome lattice. If $p<p_b$, the lattice is floppy.
Above $p_b$, there are three regions: critical (I), dense (III), and transition (II).
Each of these is
divided into a bending dominated regime (clear) at low $\kappat$ and stretching
dominated one (dotted) at high $\kappat$. $\Delta p = (p-p_b)$.}
\label{fig:phasedia}
\end{figure}

By stacking and connecting kagome lattices, we construct a 3d
model lattice with exact four-fold coordination that supports
shear and compressional stress even in the absence of bending
forces. Using analytical theory and numerical simulations, we
study the elastic properties of this lattice as a function of
the unitless measure $\kappat = \kappa/(\mu a^2)$,  where $a$
is a length scale, of the relative strength of bending compared
to stretching forces. We demonstrate analytically that our
undiluted lattice does exhibit affine, bending-independent
elastic moduli of order $\muk/a^2$ with that for pure
shear in the $xy$ plane equal to $G_0 = 9\mu/a^2$. We use
numerical simulations to study the elastic properties of our
lattice when polymers are shortened by cutting bonds with
probability $1-p$. Figure \ref{fig:phasedia} provides a phase
diagram summarizing our results. For all $p_b<p_1$ stretching
(bending) dominates response at large (small) $\kappat$. Near
the rigidity percolation threshold $p_b$, $G \sim (p-p_p)^f$
with $f \approx 0.2$ with amplitude proportional to
$G_0$ for $\kappat \gg 1$ and to $\kappa/a^4$ for
$\kappat \ll 1$. Near the dense limit $p=1$, $G /G_0$
is well described by a critical-like scaling function of $\tau
\sim \kappat/(1-p)^2 \sim \kappat L^2$, where $L = a(1-p)^{-2}$
is the average polymer length, in which $G$ approaches
$(\kappa/a^6) L^2$ for $\tau \ll 1$, but other scaling
variables cannot be ruled out. There is a transition region
(II in Fig.~\ref{fig:phasedia}) that interpolates smoothly
between between the rigidity percolation (I in
Fig.~\ref{fig:phasedia}) and the dense (III in
Fig.~\ref{fig:phasedia}) regions. These results are
consistent with those found by Broedersz, Sheinman,
and MacKintosh \cite{BroederszMac2011} in a phantom lattice
model, which like ours has a maximum coordination of $4$
and $\kappa$-independent affine moduli at $p=1$. When
$\kappa =0$, our simulations support the expected existence of
a first-order jump in $G$ at $p=1$.

We construct our 3d model lattice as illustrated in
Fig.~\ref{fig:construction}. We start with a conventional kagome
lattice with lattice constant $a$ lying in the $x$-$y$ plane with the
$x$-axis along one of the straight lines of bonds.  The kagome lattice
has 3 sites in its unit cell as indicated by the shaded triangles in
Fig.~\ref{fig:construction}. Next, we generate a stack of kagome
lattices by placing replicas of the original lattice in planes normal
to the $z$-axis with the spacing between consecutive planes equal to
$a$. Then we generate 4 replicas of this stack. The first
replica remains in its place. The second replica is rotated by an angle
$\theta = \pi/2$ about the $y$-axis and then translated by $a (1/8, 0,
1/4)$. The third replica is rotated by an angle $\phi =  \pi/3$ about
the $x$-axis and then translated by $a (0, \sqrt{3}/8, 0)$. The fourth
replica is rotated by an angle $\phi = -\pi/3$ about the $x$-axis and
then translated by $a (3/16, \sqrt{3}/8, 0)$. Finally, new lattice
sites (crosslinks) are introduced at the crossing points of polymers.
The resulting structure is shown in Fig.~\ref{fig:construction} (d). It
is a periodic lattice with strict 4-fold coordination and a tetragonal
unit cell in the form of a rectangular parallelepiped
containing $54$ sites. As in Mikado models, the segment lengths in our
model vary (from $a/2$ to $a/16$).

The elastic energy density $E$ of a filamentous network is the sum of a
stretching contribution $E_{\text{s}}$ and a bending contribution
$E_{\text{b}}$, each of which is a function of the displacements
$\brm{u}_\alpha$ of lattice crosslinks $\alpha$ from their rest
positions $\brm{x}_{\alpha}$. To harmonic order in elastic
displacement, we can write these contributions for our model lattice as
\begin{align}
\label{Estretch}
E_{\text{s}}  &= \frac{\muk}{2} \sum_{\langle \alpha, \beta \rangle} \frac{\left[  \brm{e}_{\langle \alpha, \beta \rangle} \cdot \Delta \brm{u}_{\alpha \beta}\right]^2}{\left| \Delta \brm{x}_{\alpha \beta} \right|},
\\
E_{\text{b}}  &= \sum_{\langle \gamma, \alpha, \beta\rangle} \frac{\kappa_{ \gamma \alpha\beta}}{2}
 \left[  \brm{e}_{\langle \alpha, \beta \rangle} \times \left( \frac{\Delta \brm{u}_{\alpha \beta}}{\left| \Delta \brm{x}_{\alpha \beta}  \right|} - \frac{\Delta \brm{u}_{\gamma \alpha}}{\left| \Delta \brm{x}_{\gamma \alpha}  \right|} \right) \right]^2
\end{align}
where $\Delta \brm{u}_{\alpha \beta} =  \brm{u}_\alpha - \brm{u}_\beta$ and $\Delta \brm{x}_{\alpha \beta} = \brm{x}_\alpha - \brm{x}_\beta$. $\brm{e}_{\langle \alpha, \beta \rangle}$ is
the unit vector directed from site $\alpha$ to $\beta$ in the reference
lattice.The summations run over all bonds and bond-pairs, respectively. $\kappa_{ \gamma \alpha\beta} =  2 \kappa/(\left| \Delta \brm{x}_{\alpha \beta}  \right| + \left| \Delta \brm{x}_{\gamma \alpha}  \right|)$ is a segment-length dependent bending constant derived from the worm-like-chain model. In the following, we treat $\muk$ and $\kappa$ as independent mechanical parameters; in real systems $\muk$ is dominated by entropic
stretching and is a function of temperature and $\kappa$
\cite{MacKintoshJan1995}.

Under imposed external strain, individual lattice sites undergo
displacements $u_{\alpha,i}= \eta_{ij} x_{\alpha,i} + \delta
u_{\alpha,i}$ for $i=x,y,z$, where $\eta_{ij}$ is the imposed
macroscpic deformation. When the equilibrium value of $\delta
\brm{u}_{\alpha}$ is zero, each displacement follows the macroscopic
strain, and response is {\em affine};  when it is nonzero, response is
{\em nonaffine}. In both the affine and nonaffine cases, the elastic
energy $\mathcal{E} = E/(Na^3)$ per crosslink per volume
$a^3$ of our model obtains the form appropriate to tetragonal symmetry
with $6$ independent elastic constants:
\begin{align}
\label{tetragonalEn}
\mathcal{E}&= \textstyle{\frac{1}{2}} \, C_{xxxx}\, u_{xx}^2 + \textstyle{\frac{1}{2}} \, C_{yyyy}\, \left[ u_{yy}^2   +  u_{zz}^2 \right]
\nonumber\\
&+  \textstyle{\frac{1}{2}} \, C_{xyxy} \, \left[  u_{xy}^2 +  u_{xz}^2 \right]  +  \textstyle{\frac{1}{2}} \, C_{yzyz} \, u_{yz}^2
\nonumber \\
&+ C_{xxyy} \, u_{xx} \left[ u_{yy} + u_{zz}  \right]+ C_{yyzz} \, u_{yy} u_{zz} \, ,
\end{align}
where $u_{ij} = \textstyle{\frac{1}{2}} (\eta_{ij} + \eta_{ji})$ is the
usual, linearized symmetric Lagrange strain tensor~\cite{Landau-elas}
and $N$ is the number of sites in the lattice.

We consider first the undiluted model with $\kappa=0$.  In this case, a
straightforward symbolic solution for all $\brm{u}_\alpha$ in terms of
$\eta_{ij}$ yields $\delta \brm{u}_\alpha = 0$ for all $\alpha$, i.e.,
response is purely affine.  This implies that all filaments remain
straight under elastic distortion, and, as a result, the elastic energy
is independent of $\kappa$.  Our model thus provides a proof of
principal of the existence of 3d central-force,
subisostatic lattices with purely affine response.  The elastic
constants of the undiluted lattice in units of $\mu/a^2$ read $C_{xxxx}
= \frac{25}{2}$, $C_{yyyy} = \frac{153}{16}$, $G \equiv \, C_{xyxy} = 9$,
$C_{yzyz}=\frac{51}{4}$, $C_{xxyy}=\frac{9}{4}$, and $C_{yyzz} = 9$.

The polymers forming biological gels have finite length and they are
polydisperse. Furthermore, the topology of their networks is that of a
random solid rather than of a crystalline solid with a well defined
point group symmetry. To study the influence of randomness and network
connectivity on the elasticity of our model lattice, we dilute it by
randomly removing bonds with a given probability $1-p$. Then, we
calculate its mechanical response numerically for a range of values of
$p$, $\kappa$ and number of crosslink sites $N = 54 S$, where
$S=S_xS_yS_z$ is the number of unit cells stacked
$S_x$-times in the $x$-direction, etc. We focus on the response to
shear in the $x$-$y$-plane and set $\eta_{ij}= \gamma (\delta_{ix}
\delta_{jy} + \delta_{iy} \delta_{jx})$ with imposed strain $\gamma =
0.01$ for all deformations. We generate 100 random conformations, and
for each we calculate all displacements $\delta \brm{u}_\alpha$ by
minimizing the total energy using a conjugate gradient method. For each
$\kappat$, $p$, and $N$, we calculate $G$, the standard measure of
nonaffinity \cite{HeadMacKLevine2003,DiDonna2005} $\Gamma = \frac{1}{N}
\sum_\alpha \left(  \delta \brm{u}^0_\alpha \right)^2, $ averaged over
all configurations, and the fraction $n$ of configurations with
nonvanishing $G$.
\begin{figure}
\includegraphics[width=7.2cm]{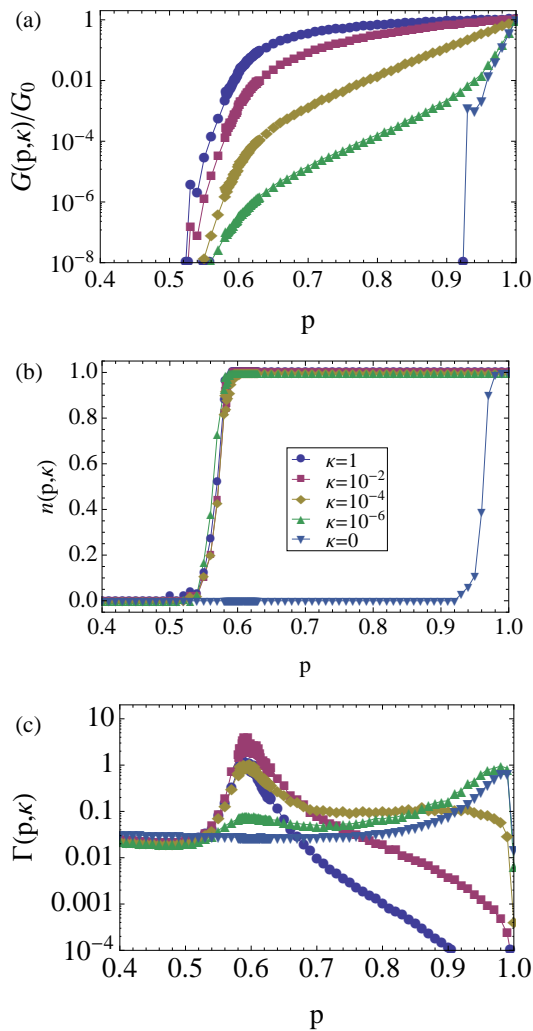}
\caption{(Color online) (a) Shear modulus $\mathcal{G}$;
(b) fraction of rigid conformations $n$; (c) non-affinity parameter $\Gamma$.}
\label{fig:combinedxyxyPlots}
\end{figure}

Figure~\ref{fig:combinedxyxyPlots} shows plots of these quantities for
$S=72$ unit cells as a function of $p$ for different $\kappat$. The
ratio $\mathcal{G}=G(p,\kappa)/G_0$  approaches $1$ as $p\to 1$, as
expected, and it does so more rapidly with increasing $\kappat$ such
that response is nearly affine for $p>0.8$ for all but the smallest
$\kappat$, indicating that response for physical dense networks can be
nearly affine. For all $\kappat>0$, $\mathcal{G}$ vanishes at a
rigidity percolation threshold $p_b (S)$. For $\kappat=0$, however,
$\mathcal{G}$ vanishes at a much larger threshold $p_c (S)$. Both $p_b
(S)$ and $p_c (S)$ decrease as the system-size decreases, with $S\to
\infty$ values of $p_c=1$ and $p_b \approx 0.602$, the latter in
good agreement with the Maxwell counting arguments of
Ref.~\cite{broedMaoLubMacK}.  The fraction $n$ reaches unity, its upper
limit, for $\kappat>0$ at $p\sim 0.6$, a value that changes little with
$S$. For smaller $p$, it drops down to zero with steepness that
increases with $S$ and that approaches a unit-step function for $S \to
\infty$. The values of $n$ between $0$ and $1$ below $p\sim  0.6$ are a
finite size effect. The corresponding conformations are the ones that
lead to nonzero values of $\mathcal{G}$ in
Fig.~\ref{fig:combinedxyxyPlots}a below $p\sim  0.6$. For $\kappa=0$,
$n$ approaches a unit step function at $p=1$ for $S \to \infty$. The
nonaffinity parameter $\Gamma$, shown in
Fig.~\ref{fig:combinedxyxyPlots}c has a peak at $p\sim  0.6$ for all
$\kappa$ and vanishes as required for $p=1$; for $p$ just below $1$,
however, it increases substantially for small $\kappat$. The location
of the peak changes little with $S$.

\begin{figure}
\includegraphics[width=7.0cm]{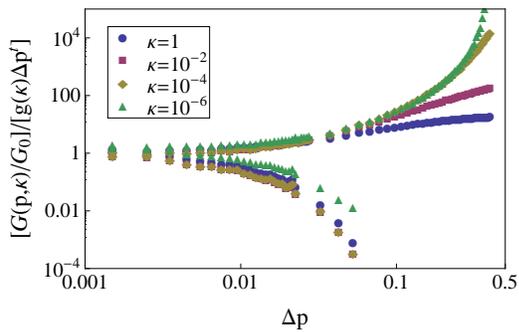}
\caption{(Color online) Scaling of $\mathcal{G}$ near $p_b$.}
\label{fig:CxyxyScalingPlot}
\end{figure}

\begin{figure}
\includegraphics[width=7.0cm]{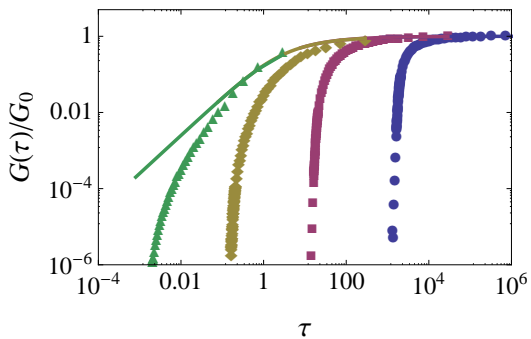}
\caption{(Color online) Scaling of $\mathcal{G}$ near $p=1$. The symbol-legend is the same as in Fig~\ref{fig:CxyxyScalingPlot}.}
\label{fig:CxyxyVsTauFit}
\end{figure}

Based on these observations, we expect the following scenario for the
infinite-size limit: For $\kappat=0$, $G$ displays a discontinuous jump
from zero to its affine value at $p=p_c =1$ reminiscent of a
first-order phase transition. For all $\kappat>0$, $G$
undergoes a rigidity percolation transition at $p_b\sim  0.6$. The
behavior for both $\kappat=0$ and $\kappat>0$ is in qualitative
agreement with the 2d kagome lattice, where the shear modulus displays
a first-order jump at $p=1$ for $\kappat=0$ and a rigidity percolation
transition at $p_b\sim  0.6$ for $\kappat>0$ \cite{MaoLub2011a}. Guided
by effective medium theory (EMT) for the 2d kagome lattice
\cite{MaoLub2011a}, we fit $\mathcal{G}$ near $p_b$ to the scaling form
\begin{align}
\mathcal{G}(p,\kappat) = g(\kappat) | \Delta p|^f,
\end{align}
where $\Delta p = p-p_b$ and $g(\kappat) = c_1 \kappat/(c_2+ \kappat)$.
As shown in Fig.~\ref{fig:CxyxyScalingPlot} data-collapse is obtained
for $p_b = 0.602$, $c_1 = 0.08$, $c_2=0.1$ and $f=0.2$, but we cannot
rule out somewhat different functions $g(\kappat)$ and values of $f$ as
small as zero, which leaves open the possibility of a weak first-order
transition. Interestingly, the exponent $f=0.2$ is not far from the
exponent $\beta = 0.175$ for the size of the percolating rigid cluster
in simulations of the transition to a rigid but unstressed state in
models of network glasses \cite{ThorpePhil2000}. Near $p=1$, the kagome
EMT suggests the scaling form
\begin{align}
\mathcal{G} (p,\kappat) = \tau^{-1} \left( -1 + \sqrt{1+\tau}\right)^2 ,
\end{align}
as long as the scaling variable $\tau = A \kappat / (1-p)^2$
is greater than $10^3 \kappat$. With $A=0.7$, this form
provides a excellent fit to our data for $\tau > 10^4 \kappat
$ as shown in Fig.~\ref{fig:CxyxyVsTauFit}. The EMT scaling form
crosses over from $G=G_0$ for $\tau \gg 1$ to $G=\frac{1}{4}G_0\tau
\sim (\kappat/a^4)(1-p)^{-2} \sim (\kappat/a^6) L^2$ for $\tau \ll 1$,
and we expect our model network to show the same crossover for $S\to
\infty$, although further simulations with larger system-sizes and
smaller values of $\kappat$ are needed to corroborate this conjecture
which is consistent with the findings for the 3d phantom
lattice~\cite{BroederszMac2011}. Over all, the EMT predictions for the
2d kagome lattice work remarkably well for our 3d kagome based lattice.
The likely explanation is that bending forces are quite effective at
restoring rigidity, and in the end, the important thing is that both
the 2d and 3d lattices have nonvanishing bulk and shear moduli at $p=1$
and $\kappa = 0$ and both exhibit a first-order rigidity transition at
$p=1$ when $\kappa = 0$.

In our network, all filaments are straight.  In real networks of
semi-flexible polymers, filaments are in general not straight, and as a
result, like the $4$-coordinated diamond lattice \cite{HeThorpe}, they
may not have nonvanishing shear and bulk moduli \cite{HuismanLub2010}
when $\kappa = 0$ even in the limit $p=1$ ($L\to \infty$).  If unit
cells in the kagome lattice are twisted through an angle
$\psi$ so that filaments are no longer straight, the lattice no longer
resists compression when $\kappa=0$, but it does resist shear
\cite{SouslovLub2010a}. These examples make it clear that network
geometry can play a roles as important as coordination number in
determining elastic response.  Further research on lattices with
different bend geometries is clearly of interest.  We have begun
studying a twisted version of our 3d lattice, and our
preliminary results indicate that at $p=1$ and $\kappa =0$ as in the
twisted kagome lattice, compression moduli vanish for all $\psi >0$ but
contrary to the kagome lattice, shear moduli remain nonzero only up to
a small critical value $\psi_c$ of $\psi$. Thus for $\psi >\psi_c$, shear
moduli at $p=1$ vanish with $\kappa$ as they do in the diamond
\cite{HeThorpe} and computer generated network lattices
\cite{Huisman1,HuismanLub2010}.

It is a pleasure to acknowledge discussions with X.\
Mao, C.P.\ Broedersz and F.C.\ MacKintosh. We also thank C.P.\ Broedersz {\em et al}.\ for sharing their manuscript~\cite{BroederszMac2011} prior to pubication.  This work was
supported by in part by NIH under grant number R01 GM083272-02
S1 (OS) and the NSF under grants DMR-0804900 and MRSEC
DMR-0520020 (TCL).

\end{document}